\documentstyle[12pt,epsf]{article}
\newcommand{\lsim}{\stackrel{<}{_\sim}}
\newcommand{\gsim}{\stackrel{>}{_\sim}}
 
\begin{document}
 
\thispagestyle{empty}
\begin{flushright}
CERN-TH/98-314\\
September 1998
 
\end{flushright}
\vspace*{1.5cm}
\centerline{\Large\bf Non-perturbative Effects in $B\to X_sl^+l^-$}
\vspace*{2cm}
\centerline{{\sc Gerhard Buchalla}}
\bigskip
\centerline{\sl Theory Division, CERN, CH-1211 Geneva 23,
                Switzerland}
 
\vspace*{1.5cm}
\centerline{\bf Abstract}
\vspace*{0.3cm}
\noindent 
The rare decay $B\to X_sl^+l^-$ provides excellent prospects for 
precision tests of Standard Model flavour dynamics. The process can be 
computed in perturbation theory with small uncertainty. However, in 
order to ensure a reliable theoretical prediction, also
potentially important effects from non-perturbative QCD have to be 
controlled with sufficient accuracy. The present article summarizes 
recent developments related to this topic.

\vspace*{3cm}
\centerline{\it Invited Talk presented at the}
\centerline{\it QCD Euroconference 98, Montpellier, 2-8 July 1998}
\vfill
 
\newpage
\pagenumbering{arabic}

\section{Introduction}

Within the Standard Model the inclusive rare $B$ decays
$B\to X_sl^+l^-$ ($l=e,$ $\mu$, $\tau$) are generated by loop-induced
flavour-changing neutral currents (FCNC). They allow us therefore
to test our understanding of flavour dynamics at the quantum level.
Although smaller in branching ratio than the radiative decay
$B\to X_s\gamma$ by almost two orders of magnitude, the $B\to X_sl^+l^-$
modes could play a decisive role because they are sensitive to additional
short-distance
contributions. $B\to X_sl^+l^-$ can thus probe aspects of flavour
physics not accessible with $B\to X_s\gamma$ and provides therefore
a very valuable, complementary source of information.
The best current limits for the (integrated, non-resonant) branching
fraction of $B\to X_s\mu^+\mu^-$ are
\begin{equation}\label{exp}
B(B\to X_s\mu^+\mu^-) <
\left\{ 
\begin{array}{l}
5.8\cdot 10^{-5}\qquad \cite{CLEO}\\
3.2\cdot 10^{-4}\qquad \cite{D0}
\end{array}
\right.
\end{equation}
This is to be compare with a Standard-Model expectation of about
$6\cdot 10^{-6}$. The modes $B\to X_sl^+l^-$ ($l=e,$ $\mu$), to which we
shall restrict ourselves in the following, should be well within reach of 
the upcoming experiments at the $B$-factories, the Tevatron and the LHC.

The theoretical description of $B\to X_sl^+l^-$ is based on a low-energy
effective Hamiltonian ${\cal H}_{eff}$, derived using the operator-product
expansion and renormalization-group improved QCD perturbation theory.
Schematically, the amplitude for the decay of $B$ into final state 
$X_sl^+l^-$ may be written as
\begin{equation}
\langle X_sl^+l^-|{\cal H}_{eff}|B\rangle =
-\frac{G_F}{\sqrt{2}}V^*_{ts}V_{tb}
\sum_i C_i(M_W/\mu,m_t,\alpha_s)\cdot
\langle X_sl^+l^-|Q_i|B\rangle
\end{equation}
The $C_i$ are Wilson coefficient functions containing the
short-distance physics from scales $M_W$ to $\mu\sim m_b$.
They are perturbatively calculable. $Q_i$ are the corresponding
local four-fermion operators. (See \cite{BBL} for a general review
on this subject.)
For the inclusive decay of a $B$ meson the squared matrix elements of
operators entering the decay rate can be further treated using the
heavy quark expansion (HQE) \cite{BBSUV,CGG}.
To leading order in this expansion in $1/m_b$ the inclusive rate
for the decay $B\to X_sl^+l^-$ is given by the partonic result for
$b\to sl^+l^-$, plus non-perturbative power corrections of second
and higher order in $\Lambda_{QCD}/m_b$. Unlike exclusive modes, the
inclusive $B\to X_sl^+l^-$ is thus largely dominated by perturbatively
calculable contributions. Non-perturbative mechanisms are, however,
still present. The ultimate precision of the Standard-Model prediction,
and the potential to test for unexpected deviations, then depends on
how well the generally difficult non-perturbative sector can be
brought under control.

\section{Overview}

\begin{figure}[htb]
 \epsfysize=7cm 
 \epsfxsize=7cm
 \centerline{\epsffile{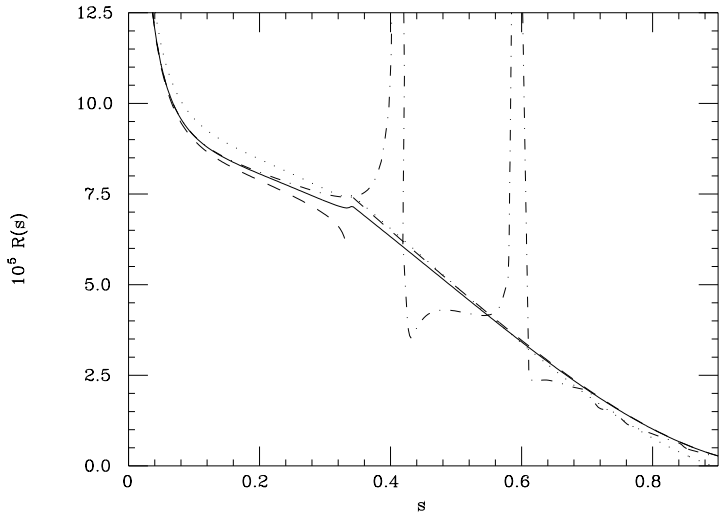}} 
\caption{Dilepton invariant mass spectrum
of $B\to X_s e^+e^-$ normalized to the semileptonic rate: 
partonic result (full line), partonic 
result + ${\cal O}(\Lambda^2_{QCD}/m^2_b)$ corrections (dotted line), 
partonic result + ${\cal O}(\Lambda^2_{QCD}/m^2_c)$  corrections
(dashed line), partonic result + factorizable resonance contributions 
(dash-dotted line). The plot has been obtained for 
$\mu=m_b=4.8$ GeV and $m_c=1.4$ GeV.}
\label{fig1}
\end{figure}

Before discussing some of the most relevant issues in slightly
more detail, we begin with a short overview of the non-perturbative
effects in $B\to X_sl^+l^-$. For illustration we show in 
Fig. \ref{fig1} the partonic dilepton invariant-mass spectrum
\begin{equation}\label{rsdef}
R(s)=\frac{\frac{d}{ds}\Gamma(B\to X_sl^+l^-)}{\Gamma(B\to X_ce\nu)}
\end{equation}
as a function of $s=q^2/m^2_b$, $q^2=(p_{l^+}+p_{l^-})^2$.
The effect of various non-perturbative corrections, to be discussed
below, is also illustrated. Clearly visible are the narrow resonances
$J/\Psi(3097)$ and $\Psi^\prime(3686)$ at $s\sim 0.4$ and $\sim 0.6$,
respectivley.

Two different sources of non-perturbative contributions may be 
distinguished. First, power corrections in $1/m_b$ to the leading
partonic decay within the heavy quark expansion. Second,
non-perturbative effects related to intermediate $c\bar c$ pairs
($B\to X_s(c\bar c)\to X_sl^+l^-$).

The leading $1/m_b$ corrections of ${\cal O}(\Lambda^2_{QCD}/m^2_b)$
have been calculated and are found to be small 
($\lsim$ several per cent), except for $q^2\gsim 0.8 m^2_b$
\cite{AHHM,BI1}. Still closer to the endpoint $q^2\approx m^2_b$
the HQE breaks down and the inclusive decay $B\to X_sl^+l^-$ 
is eventually dominated by the exclusive channel $B\to Kl^+l^-$.
(The issue of the $b$-quark mass in the partonic spectrum is
briefly discussed in \cite{BI1}. In the present context the
pole mass definition is used for $m_b$.)

Large backgrounds to the short-distance process arise from the
cascades $B\to X_s\Psi^{(\prime)}$, $\Psi^{(\prime)}\to l^+l^-$
when the dilepton mass $q^2$ is in the vicinity of the 
narrow ($J^{PC}=1^{--}$) resonances $\Psi$ and $\Psi^\prime$. 
They would dominate the
signal by two orders of magnitude and have to be excluded by suitable
cuts on $q^2$. This leaves essentially two interesting $q^2$ regions
to search for $B\to X_sl^+l^-$: region I (below the resonances), where
residual non-perturbative effects from $c\bar c$ resonances can be
calculated within a $1/m_c$ expansion; and region II (above the 
$\Psi^\prime$), where the impact of higher $J^{PC}=1^{--}$ 
$c\bar c$-resonances has to be considered.
  
\section{$\Lambda^2_{QCD}/m^2_b$ Corrections}

The starting point for a systematic expansion in $1/m_b$ 
of an inclusive $B$ meson decay rate $\Gamma$ is provided by the
optical theorem \cite{BBSUV},
\begin{equation}\label{opth}
\Gamma=\frac{1}{2M_B}\langle B|\, {\rm Im}\, i\int d^4x
T {\cal H}_{eff}(x){\cal H}_{eff}(0)|B\rangle
\end{equation}
which relates the decay width to the absorbtive part of the
forward scattering amplitude. In the present context
${\cal H}_{eff}$ is the effective hamiltonian for $b\to sl^+l^-$
decay. The hadronic part of ${\cal H}_{eff}$ involves currents
of flavour structure $(\bar bs)$ (and $(\bar sb)$).
Schematically, the T-product in (\ref{opth}) then gives rise to
expressions of the form $(\bar bs)(x) (\bar sb)(0)$, where the
$s$-quark fields are contracted according to Wick's theorem
and the $s$-quark propagator is subject to the HQE in $1/m_b$.
Through order $\Lambda^2_{QCD}/m^2_b$ the rate becomes 
\begin{equation}\label{ddsg}
\frac{d}{ds}\Gamma(B\to X_sl^+l^-)=\frac{d}{ds}\Gamma(b\to sl^+l^-)
\frac{\langle B|\bar bb|B\rangle}{2M_B}  
+C_G\frac{1}{m^2_b}
\frac{\langle B|\bar b g\sigma\cdot G b|B\rangle}{2M_B}
\end{equation}
with a calculable coefficient $C_G$ \cite{AHHM,BI1}.
The matrix elements in (\ref{ddsg}) may be further expanded using
heavy quark effective theory (HQET) \cite{MW}
\begin{eqnarray}\label{la12}
\frac{\langle B|\bar bb|B\rangle}{2M_B} &=&
  1+\frac{\lambda_1}{2m^2_b}+\frac{3}{2}\frac{\lambda_2}{m^2_b} \\
\frac{\langle B|\bar b g\sigma\cdot G b|B\rangle}{2M_B} &=&
6\lambda_2=\frac{3}{2}(M^2_{B^*}-M^2_B) \label{la2}
\end{eqnarray}
where $\lambda_1=\langle B|\bar h(iD)^2h|B\rangle/(2M_B)$
and $6\lambda_2=\langle B|\bar h g\sigma\cdot Gh|B\rangle/(2M_B)$,
with $h$ the $b$-quark field in HQET.

It is useful to normalize the rate (\ref{ddsg}) by the rate of the
inclusive, charged-current decay $B\to X_ce\nu$, defining
$R(s)$ in (\ref{rsdef}).
In $R(s)$ the CKM parameters enter only in the ratio
$|V_{ts}V_{tb}/V_{cb}|^2=0.95\pm 0.03$, which is much better known
than the individual elements. In addition the dependence on the
poorly known parameter $\lambda_1$ drops out in the
$1/m^2_b$ corrections to $R$. These are then determined by $\lambda_2$
alone, which is fixed by (\ref{la2}).
The relative correction $\delta_{1/m^2_b}R(s)/R(s)$ amounts to
several per cent at most for $s\lsim 0.75$, but diverges towards
the endpoint $s\to 1$. This breakdown of the HQE was first noticed
in \cite{AHHM} and confirmed in \cite{BI1}. The origin of this behaviour
has been discussed in detail in \cite{BI1}. The reason for the 
breakdown of the HQE can be understood intuitively as follows.
At $s\approx 1$ the lepton pair carries almost the entire mass of the
initial $B$ meson, leaving a very soft, low-multiplicity hadronic
system in the final state. No hard scale exists in this regime to
define an expansion parameter for the strange-quark propagator.
It is interesting to note \cite{BI1} that the nature of the
HQE breakdown in $B\to X_sl^+l^-$ at $s\approx 1$ is conceptually
different from the situation of the lepton (photon) energy spectrum
in $B\to X_ce\nu$ ($B\to X_s\gamma$) \cite{BSUV}-\cite{NEU12}.
In the latter cases the straigtforward HQE breaks down near the endpoint
of the lepton (photon) energy spectrum, but the expansion can be amended
by a resummation of the leading singular contributions to all
orders in $1/m_b$. The resummation results in a description of the endpoint
region in terms of a shape function \cite{BSUV}-\cite{NEU12}.
No such resummation is possible near the $s\to 1$ endpoint of 
$B\to X_sl^+l^-$ and the shape function description does not apply
\cite{BI1}.
On the other hand, the kinematic situation at $s\approx 1$ is a priori
very well suited for a treatment employing heavy hadron chiral
perturbation theory \cite{WI,BD}. This framework is appropriate
to analyze the strong-interaction effects in the transition of
a heavy $B$ meson to a small number of soft pseudo-goldstone bosons.
This is precisely the case of $B\to X_sl^+l^-$ at $s\approx 1$, where
the inclusive decay degenerates into $B\to Kl^+l^-$,
$B\to K\pi l^+l^-$, $\ldots$, with low-energy kaons and pions.
This approach to the endpoint region has been discussed in 
\cite{BI1,FG}. $B\to Kl^+l^-$ largely domiantes the endpoint region,
whereas $B\to K\pi l^+l^-$ is entirely negligible. A reasonable 
interpolation to the range of validity of the HQE at lower $s$
can be obtained \cite{BI1}.

\section{$\Lambda^2_{QCD}/m^2_c$ Corrections} 

For small $q^2\lsim 3m^2_c$ (that is below the $\Psi$ resonance),
the leading non-perturbative effect from intermediate $c\bar c$
can be treated in an expansion in $1/m_c$. This approach has first
been proposed to estimate non-perturbative corrections in
$B\to X_s\gamma$ \cite{VOL}. The generalization to $B\to X_sl^+l^-$
was performed in \cite{BIR} (see also \cite{CRS}). The effect
arises from the interaction of a soft gluon coupled to the
$c\bar c$ loop with the spectator cloud inside the $B$ meson.
This mechanism is not included in the perturbative evaluation of the
$B\to X_sl^+l^-$ decay rate. The leading correction has the form
$\Delta\Gamma(B\to X_sl^+l^-)=f(s)\cdot \lambda_2/m^2_c$,
where $f(s)$ is a function of the dilepton mass, $m_c/m_b$ and the
short-distance Wilson coefficients. Since the hadronic matrix element
can be expressed entirely in terms of the well known $\lambda_2$
(\ref{la2}), the $1/m^2_c$ effect is completely calculable. A further
virtue of this approach is its model independence. In particular it
avoids the double counting of $c\bar c$ contributions 
(already partly contained in the perturbative calculation) that is
inherent in resonance-exchange models that have been employed 
previously to estimate these long-distance effects.

Numerically the $1/m^2_c$ correction is below a few per cent for
$q^2\lsim 3m^2_c$. The $1/m_c$ expansion is not valid in the vicinity of
$q^2=M^2_\Psi, M^2_{\Psi'}$, where strong resonance effects are
important. For higher values of $q^2$ the $1/m^2_c$ correction
becomes very small. However, in this regime the non-perturbative
contributions are dominated by the effect of higher $c\bar c$ resonances,
which will be briefly discussed in the following section.

To conclude we remark that the $1/m^2_c$ correction to
$B\to X_s\gamma$ \cite{VOL,BIR,LRWG} may be obtained from the
corresponding calculation for $B\to X_sl^+l^-$ by considering the
limit $q^2\to 0$. One finds
$\Delta\Gamma(B\to X_s\gamma)/\Gamma(B\to X_s\gamma)=$
$0.42\, \lambda_2/m^2_c\approx +3\%$ \cite{BIR}.

\section{Higher $c\bar c$-Resonances}

In the region of $q^2$ above the $\Psi^\prime(3686)$ there are further 
resonances ($\Psi(3770)$, $\Psi(4040)$, $\Psi(4160)$, $\Psi(4415)$), 
which may affect the dilepton invariant mass spectrum. Since these 
higher resonances are rather broad and much less prominent than 
$\Psi$ and $\Psi^\prime$, the HQE-based quark-level calculations 
should give a reasonable first approximation in the corresponding range of
$q^2$. To calculate the impact of higher resonances on the spectrum above
$q^2=M^2_{\Psi^\prime}$ is a difficult problem, which has not yet been 
solved in a fully satisfactory way. Some insight can be gained by making 
further simplifying, model-dependent assumptions.
A procedure that is often employed in the literature \cite{AHHM} to 
estimate the resonance contributions consists in adding Breit-Wigner 
type resonance terms to the quark-level calculation. This approach is, 
however, unsatisfactory due to the double-counting problem
mentioned above. At present, the most consistent method is the one 
proposed in \cite{KS}. Here the charm-loop contribution to $B\to X_sl^+l^-$ 
is estimated by means of experimental data on 
$\sigma(e^+e^-\to c\bar c-{\rm hadrons})$ using a dispersion relation.
In this way double counting is avoided. The estimate has the further 
advantage of also including open-charm contributions. On the other hand, 
it still rests on the ad-hoc assumption that the
$B\to X_s c\bar c$ transition can be factorized into a product of 
$(\bar sb)$ and $(\bar cc)$ color-singlet currents. The long-distance 
corrections to the quark-level calculation estimated with this
method are only several per cent \cite{BI1}. 
Somewhat larger effects are obtained when a phenomenological
factor $\kappa=2.3$ is introduced to correct for the factorization 
approximation. This is motivated
by the fact that the factorization assumption gives a too small 
$B\to X_s\Psi$ branching fraction.
However, the validity of such a procedure for estimating the impact of 
higher resonances in $B\to X_sl^+l^-$ is not a priori clear and 
requires further study.

\section{Summary}

After large backgrounds from $B\to X_s\Psi(\Psi^\prime)$ 
$\to X_sl^+l^-$ are eliminated by
experimental cuts in $q^2$, the rare decay $B\to X_sl^+l^-$ is
dominated by perturbatively calculable contributions.
${\cal O}(\Lambda^2_{QCD}/m^2_b)$ non-perturbative corrections are
well under control, except for the endpoint region $q^2\to m^2_b$,
where the HQE breaks down. Close to the endpoint the inclusive decay
is determined by $B\to Kl^+l^-$ and model independent constraints from
heavy-hadron chiral perturbation theory can be derived.

Non-perturbative effects further arise from intermediate $c\bar c$ 
states in $B\to X_sl^+l^-$. At low $q^2$ they can be treated with a
$1/m_c$ expansion. The leading correction of 
${\cal O}(\Lambda^2_{QCD}/m^2_c)$ has been calculated and is found
to be very well under control ($\lsim$ few per cent). The corresponding
effect is even smaller at high $q^2$. There, however, higher resonances
(above $\Psi'$) can have some impact on the dilepton-mass spectrum.
These resonances are rather broad and their effect on the spectrum
is possibly not too pronounced. This question cannot yet been answered
in a fully satisfactory way and deserves further study.
Otherwise, in particular in the low-$q^2$ region, the Standard-Model
prediction is quite accurate ($\lsim 10\%$ uncertainty).
The inclusive rare decay $B\to X_sl^+l^-$ is therefore very well suited
for detailed tests of the underlying flavour physics.

\section*{Acknowledgements}

I am grateful to my co-authors in \cite{BI1,BIR} for a most enjoyable
collaboration on the topics presented in this talk. Particular thanks
are due to Gino Isidori for the extensive and fruitful exchange of ideas
on non-perturbative effects in $B\to X_sl^+l^-$.

\vfill\eject


\begin{thebibliography}{99}
\bibitem{CLEO}
S. Glenn et al. (CLEO collaboration), Phys. Rev. Lett. 80 (1998) 2289.
\bibitem{D0}
B. Abbott et al. (D0 collaboration), Phys. Lett. B423 (1998) 419.
\bibitem{BBL}
G. Buchalla, A.J. Buras and M.E. Lautenbacher,
Rev. Mod. Phys. 68 (1996) 1125.
\bibitem{BBSUV}
I.I. Bigi et al., in B-Decays (2nd edition), ed. S.L. Stone,
World Scientific, Singapore (1994), hep-ph/9401298.
\bibitem{CGG}
J. Chay, H. Georgi and B. Grinstein, Phys. Lett. B247 (1990) 399.
\bibitem{AHHM}
A. Ali, G. Hiller, L.T. Handoko and T. Morozumi, 
Phys. Rev. D55 (1997) 4105.
\bibitem{BI1}
G. Buchalla and G. Isidori, Nucl. Phys. B525 (1998) 333.
\bibitem{MW}
A.V. Manohar and M.B. Wise, Phys. Rev. D49, (1994) 1310.
\bibitem{BSUV}
I.I. Bigi, M.A. Shifman, N.G. Uraltsev and A.I. Vainshtein,
Int. J. Mod. Phys. A9, (1994) 2467.
\bibitem{MN}
T. Mannel and M. Neubert, Phys. Rev. D50, (1994) 2037.
\bibitem{NEU12}
M. Neubert, Phys. Rev. D49, (1994) 3392 and 4623.
\bibitem{WI}
M.B. Wise, Phys. Rev. D45, (1992) R2188.
\bibitem{BD}
G. Burdman and J.F. Donoghue, Phys. Lett. B280, (1992) 287.
\bibitem{FG}
A.F. Falk and B. Grinstein, Nucl. Phys. B416, (1994) 771.
\bibitem{VOL}
M.B. Voloshin, Phys. Lett. B397, (1997) 275;
A. Khodjamirian et al.,
Phys. Lett. B402, (1997) 167.
\bibitem{BIR} 
G. Buchalla, G. Isidori and S.J. Rey, Nucl. Phys. B511, (1998) 594.
\bibitem{CRS} 
J.-W. Chen, G. Rupak and M.J. Savage, Phys. Lett. B410, (1997) 285.
\bibitem{LRWG}
Z. Ligeti, L. Randall and M.B. Wise, Phys. Lett. B402 (1997) 178;
A.K. Grant et al., Phys. Rev. D56 (1997) 3151.
\bibitem{KS}
F. Kr\"{u}ger and L.M. Sehgal, Phys. Lett. {\bf B380}, 199 (1996).
\end{thebibliography}
\end{document}